# Geometric Structure of Pseudo-plane Quadratic Flows

Che Sun[1,2]

[1] Institute of Oceanology, Chinese Academy of Sciences, Qingdao, China

[2] National Laboratory for Marine Science and Technology, Qingdao, China

(Corresponding email: csun@qdio.ac.cn)

**Abstract.** Quadratic flows have the unique property of uniform strain and are commonly used in turbulence modeling and hydrodynamic analysis. While previous application focused on two-dimensional homogeneous fluid, this study examines the geometric structure of three-dimensional quadratic flows in stratified fluid by solving a steady-state pseudo-plane flow model. The complete set of exact solutions reveals that steady quadratic flows have invariant conic type in non-rotating frame and non-rotatory vertical structure in rotating frame. Three baroclinic solutions with vertically non-aligned structure disprove an earlier conjecture. The rich topology of quadratic flows stands in contrast to the depleted geometry of high-degree polynomial flows. A paradox in the steady solutions of shallow-water reduced-gravity models is also explained.

# I. Introduction

Quadratic flow is a special type of planar flows in which velocity components vary linearly with spatial coordinates. The velocity gradient and strain rate are spatially uniform. Due to this unique property, quadratic flow has been commonly used as mean flow in turbulence model and hydrodynamic stability analysis (*Lagnado et al.* 1984, *Craik and Criminale* 1986, *Salhi et al.* 1996, *Shapiro and Fedorovich* 2012). Such quadratic basic state is an exact solution of the Euler equations because viscous terms are identically zero.

Quadratic flow is also important for vortex dynamics because uniform strain field represents the distortional influence of distant vortices (*Moore and Saffman* 1971, *Kida* 1981, *Lin and Crocos* 1984, *Dritschel* 1990, *Trieling et al.* 1997). Exact vortex solutions in quadratic form have been obtained for shallow-water reduced-gravity models (*Cushman-Roisin et al.* 1985, *Young* 1986, *Ruddick* 1987, *Rubio and Dotsenko* 2006). Though nonlinear partial differential equations from these models are generally insoluble, they reduce to tractable ordinary differential equations when polynomial solutions are considered.

Previous analyses of quadratic flow are mainly for two-dimensional homogeneous fluid. The flow structure becomes considerably complex in three-dimensional non-homogeneous fluid. This study focuses on a quasi-three-dimensional type of stratified flow called pseudo-plane flow, which has vertically varying horizontal velocities but no vertical velocity (*Saccomandi* 1994). The goal is to find the complete set of exact quadratic solutions to the steady-state Euler equations represented by a pseudo-plane ideal flow (PIF) model. The result allows us to compare with the high-degree polynomial solutions obtained in *Sun* (2016). By extending to stratified fluid, we also hope to find the baroclinic counterpart of the two-dimensional Kirchhoff elliptic vortex.

## II. Pseudo-plane ideal flow

The PIF model proposed by *Sun* (2008) is essentially the steady-state Euler equations for rotating stratified fluid under Boussinesq approximation:

$$uu_x + vu_y - fv = -p_x \qquad (1)$$

$$uv_x + vv_y + fu = -p_y \qquad (2)$$

$$\rho = -p_z \qquad (3)$$

$$u_x + v_y = 0 \qquad (4)$$

$$u\rho_x + v\rho_y = 0 \qquad (5)$$

where $p$ is pressure perturbation divided by a mean density $\rho_0$, $\rho$ is density perturbation scaled by $\rho_0/g$, and $f$ is the constant Coriolis parameter (f-plane assumption). The pseudo-plane velocities are $\mathbf{u} = [u(x,y,z), v(x,y,z), 0]$. Background pressure $\bar{p}(z)$ and density $\bar{\rho}(z)$ are neglected in hydrostatic balance (3) and pressure perturbation, if exist, must have horizontal variations.

The incompressible condition (4) yields a streamfunction $\psi$ satisfying $u = -\psi_y$, $v = \psi_x$. While density conservation (5) serves as a compatibility condition for pseudo-plane flows, momentum equations (1-2) produce another compatibility condition in the form of vorticity equation $J(\psi, \zeta) = 0$, where vertical vorticity $\zeta = v_x - u_y = \nabla^2 \psi$. This vorticity compatibility condition is redundant for quadratic flows.

A steady pseudo-plane flow is equivalent-barotropic (EB) if its horizontal velocity vector does not change direction vertically and satisfies $J(\psi, \psi_z) = 0$ everywhere. Unlike the EB concept in meteorology, the definition here is purely a geometric characterization of vertical alignment and is not related to geostrophic dynamics.

A pseudo-plane flow is baroclinic if isobaric surfaces and isopycnal surfaces do not coincide. If isopycnals are flat ($\rho = 0$), the flow is degenerate and essentially a solution for homogeneous fluid. Isobaric surfaces in a baroclinic flow can not be flat, otherwise $p_x = p_y = 0$ and vertical differentiation of Eq.(3) would give $\rho_x = \rho_y = 0$.

Because baroclinic flows always have horizontally varying pressure, the geometric properties in *Sun* (2008) can be rephrased to apply to general pseudo-plane flows

:

**Theorem 1**. *If a steady pseudo-plane flow is EB, it belongs to constant-speed flow and appears as straightline jet or circular vortex.*

**Theorem 2**. *A steady pseudo-plane flow with straightline or circular streamlines must be EB if pressure perturbation exists.*

The only known baroclinic PIF solutions at the time of Sun08 were straightline jet and circular vortex, prompting a conjecture as follows: baroclinic solutions to the PIF model are always EB. While the high-degree ($n > 2$) polynomial solutions obtained by *Sun* (2016) appear to support the conjecture, we will see it breaks down in quadratic flows.

## III. Quadratic flows

We set to solve the PIF model analytically to obtain exact solutions for quadratic flows. Pseudo-plane streamfunction in general quadratic form is

$$\psi(x, y, z) = a_1(z)x^2 + a_2(z)xy + a_3(z)y^2 + b_1(z)x + b_2(z)y \qquad (6)$$

It represents elliptical flow if $\Delta = 4a_1 a_3 - a_2^2 > 0$, hyperbolic flow if $\Delta < 0$, and parabolic flow if $\Delta = 0$. At each depth we can eliminate the $xy$ term by rotating coordinate $(x, y)$ to $(\hat{x}, \hat{y})$ via

$$x = \hat{x}\cos\alpha - \hat{y}\sin\alpha, \quad y = \hat{x}\sin\alpha + \hat{y}\cos\alpha$$

where
$$\alpha = \frac{\pi}{4} \text{ if } a_1 = a_3,$$
$$\alpha = \frac{1}{2}\arctan(\frac{a_2}{a_1 - a_3}) \text{ if } a_1 \neq a_3 \qquad (7)$$

In the new coordinates $\psi$ becomes

$$\hat{\psi} = \frac{1}{2}(a_1 + a_3 + A)\hat{x}^2 + \frac{1}{2}(a_1 + a_3 - A)\hat{y}^2, \tag{8}$$
$$A^2 = (a_1 - a_3)^2 + a_2^2$$

Quadratic flow (6) always satisfies the vorticity compatibility condition because its vertical vorticity is horizontally uniform:

$$u = -a_2 x - 2a_3 y - b_2, \quad v = 2a_1 x + a_2 y + b_1$$
$$\zeta = v_x - u_y = 2(a_1 + a_3)$$

Strain rate includes two parts, namely volumetric strain rate and shear strain rate. An incompressible flow has zero volumetric strain rate, i.e., $u_x + v_y = 0$. For the quadratic flow (6), its shear strain rate is $v_x + u_y = 2(a_1 - a_3)$. Therefore circular vortex has zero shear strain rate. Elliptic vortex has small shear strain rate and belongs to weak flow. Hyperbolic flow has large shear strain rate and belongs to strong flow where strain rate is greater than vorticity.

Substituting (6) into density equation (5) yields

$$2r_1 x^2 + 2r_2 y^2 + 4r_3 xy + r_4 x + r_5 y + r_6 = 0$$

$$\begin{aligned}
r_1 &= -a_2 C + (a_1 a_2' - a_2 a_1')f \\
r_2 &= a_2 C + (a_2 a_3' - a_3 a_2')f \\
r_3 &= (a_1 - a_3)C + (a_1 a_3' - a_3 a_1')f \\
r_4 &= 2a_1 A + a_2 B - 2b_2 C + (2a_1 b_2' - 2a_1' b_2 + a_2' b_1 - a_2 b_1')f \\
r_5 &= a_2 A + 2a_3 B + 2b_1 C + (2a_3' b_1 - 2a_3 b_1' + a_2 b_2' - a_2' b_2)f \\
r_6 &= b_1 A + b_2 B + (b_1 b_2' - b_1' b_2)f \\
A &= (2a_1 b_2 - a_2 b_1)', \quad B = (a_2 b_2 - 2a_3 b_1)', \quad C = 2(a_1 a_3)' - a_2 a_2'
\end{aligned} \tag{9}$$

For exact solutions, the coefficients in (9) shall be zero everywhere, i.e., $r_1 = r_2 = r_3 = r_4 = r_5 = r_6 = 0$. The prime sign denotes vertical differentiation with respect to *z*.

## 1) Non-rotating frame

We start with a property of non-rotating quadratic flows with proof given in Appendix A.

**Lemma 1 (Quadratic flow).** *In non-rotating frame the conic type of a quadratic flow is vertically invariant.*

It means that the type of a quadratic flow, including ellipse, hyperbola and parabola, remains the same vertically. For example, if at one depth the quadratic flow is elliptical, it shall be elliptical at other depths.

Solving (9) in non-rotating frame yields a series of quadratic solutions as listed below. In parenthesis is the solution number from Appendix A.

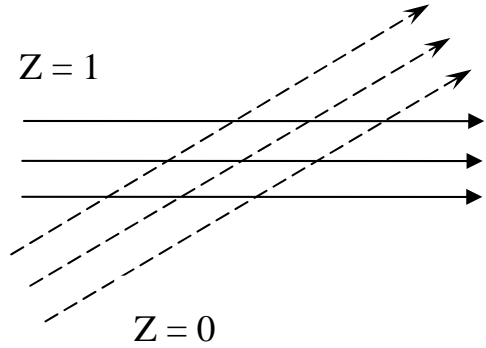

**Figure 1**. Straightline jet with direction varying vertically (S1).

**(S1)** Straightline jet (A11)

$$\psi = [a(z)x + b(z)y + h(z)]^2$$
$$u = -2b(ax+by+h), \quad v = 2a(ax+by+h)$$
$$p = 0, \quad \rho = 0$$

The straightline jet is generally non-EB as its direction varies with depth (Figure 1). The non-EB formulation does not violate Theorem 2

because pressure perturbation is absent. The jet becomes EB when ratio $b/a$ is constant or one of $a$ and $b$ is zero.

**(S2)** Circular vortex (A3)

$$\psi = a(z)(x^2 + y^2)$$
$$u = -2ay, \quad v = 2ax$$
$$p = 2a^2(x^2 + y^2), \quad \rho = -4aa'(x^2 + y^2)$$

The solution is a baroclinic EB circular vortex. In non-rotating frame, centrifugal acceleration can only be balanced by horizontal pressure gradient. The presence of pressure perturbation requires circular vortex to be EB according to Theorem 2. It also shows that baroclinicity is not unique to rotating frame and can be generated by circular vortex in non-rotating fluid.

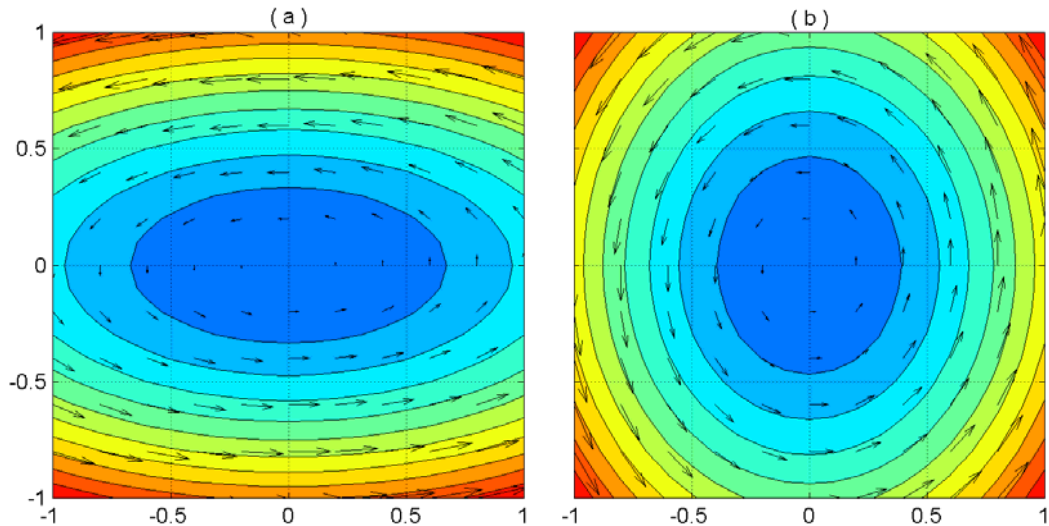

**Figure 2.** Streamfunction and velocity fields for elliptic flow of solution (S3) at (a) $z = 0.5$ and (b) $z = 1.2$, with $a = z$, $c = 1$.

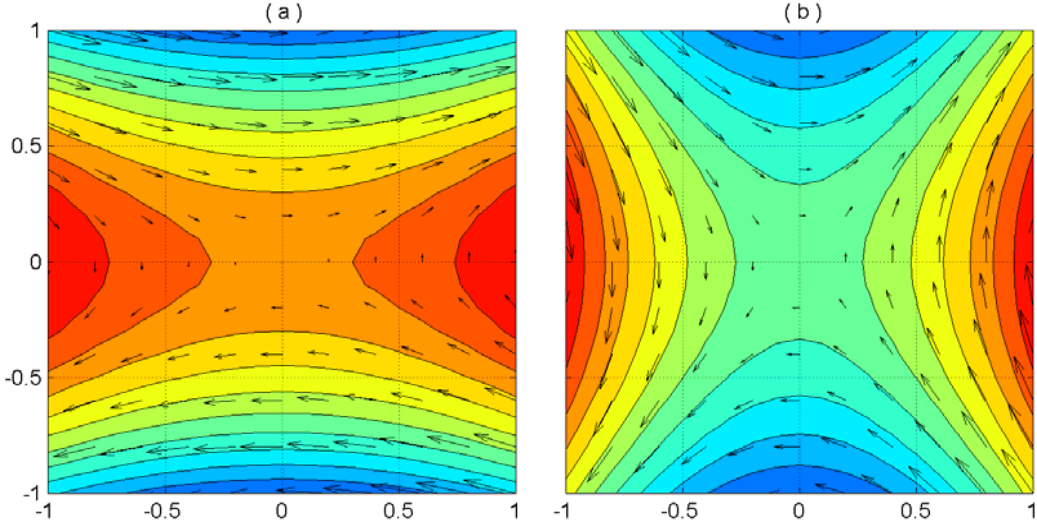

**Figure 3.** Streamfunction and velocity fields for hyperbolic flows of solution (S3) at (a) $z = 0.5$ and (b) $z = 1.5$, with $a = z$, $c = 1$.

**(S3)** Elliptical or hyperbolic flow in pulsation mode (A5)

$$\psi = a(z)x^2 + \frac{c}{a(z)} y^2$$

$$u = -\frac{2c}{a} y, \quad v = 2a\, x$$

$$p = 2c(x^2 + y^2), \quad \rho = 0$$

As $\Delta = 4c$, the flow is elliptical if $c > 0$ (Figure 2), and hyperbolic if $c < 0$ (Figure 3). The symmetry axes of the flow do not rotate with depth. For the elliptical case, the vortex pulsates as its eccentricity varying with depth, but the area enclosed by a streamfunction of the same value is constant as $c\pi\psi^{-2}$.

To further examine its geometry, we calculate

$$J(\psi, \psi_z) = uv_z - u_z v = -\frac{8ca'}{a} xy$$

$$J(\psi, K) = 8c(a - \frac{c}{a})xy$$

The solution is not an EB flow because $J(\psi,\psi_z)=0$ does not hold everywhere. It does have vertically unidirectional velocity along the x-axis and y-axis, since $J(\psi,\psi_z)=0$ at $x=0$ and $y=0$. The solution is not a constant-speed flow either, because $J(\psi,K)=0$ is only valid on the x-axis and y-axis (kinetic energy $K=(u^2+v^2)/2$). This example demonstrates that the concepts of EB flow and constant-speed flow describe global flow properties and are not defined at local point.

**(S4)** Parabolic flow (A4)

$$\psi = a(z)[x+x_0(z)]^2 + \frac{cy}{a(z)}$$
$$u = -c/a, \quad v = 2a(x+x_0)$$
$$p = 2cy, \quad \rho = 0$$

As shown in Figure 4, its symmetry line $x=-x_0$ shifts horizontally and does not rotate with depth.

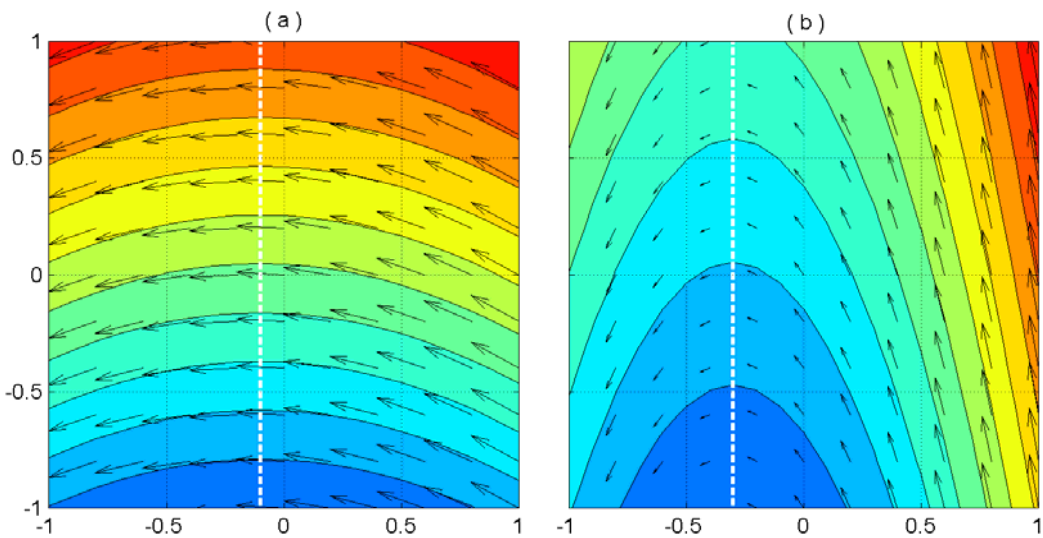

**Figure 4**. Streamfunction and velocity fields for parabolic solution (S4) at (a) $z=0.5$ and (b) $z=1.5$, with $a=z$, $x_0=0.2z$.

**(S5)** Hyperbolic flow in rotary mode (A14)

In non-rotating frame hyperbolic solutions such as (A13) can rotate and deform at the same time. Here we examine the shape-preserving hyperbolic solution (A14):

$$\psi = (c_1 \cos^2 Z - c_2 \sin^2 Z) x^2 + (c_1 \sin^2 Z - c_2 \cos^2 Z) y^2$$
$$\pm (c_1 + c_2) xy \sin 2Z$$
$$p = -2c_1 c_2 (x^2 + y^2), \quad \rho = 0$$
$$Z = Z(z), \quad c_1 > 0, \ c_2 > 0$$

Based on formulae (7) and (8), the flow field at each depth can rotate by $\alpha = \pm Z(z)$ to become the same canonical form $\hat{\psi} = c_1 \hat{x}^2 - c_2 \hat{y}^2$. It proves the shape-preserving character of this hyperbolic flow (Figure 5).

If $c_1 = c_2 = a$, the flow becomes $\psi = a(x^2 - y^2)\cos 2Z \pm 2axy \sin 2Z$ with azimuthal symmetry. It can be converted into cylindrical coordinates as $\psi = a r^2 \cos(2Z \pm 2\theta)$, which is a 3D version of the dipolar strain field frequently used in vortex studies (*Trieling et al.* 1997).

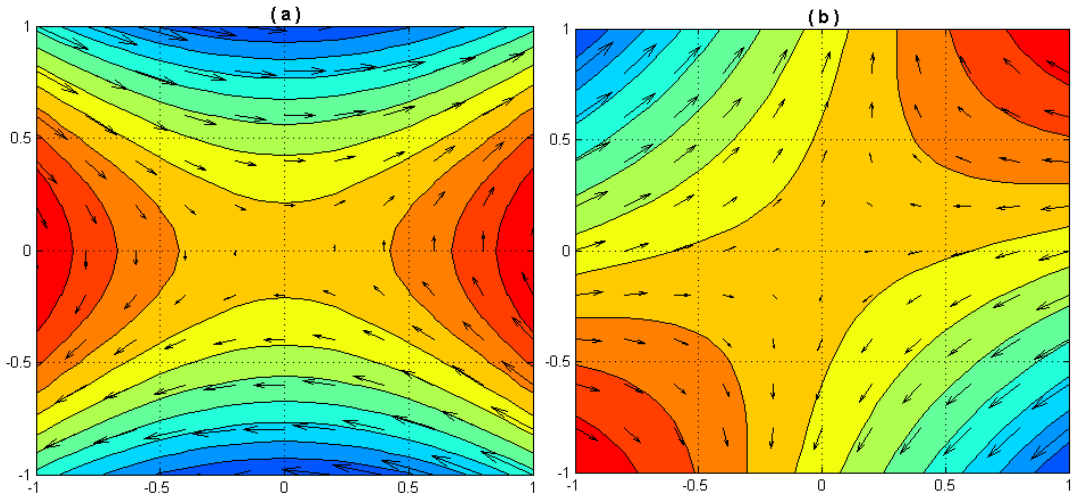

**Figure 5**. Streamfunction and velocity vectors for dipolar strain-flow solution (S5) at (a) $Z = 0$ and (b) $Z = \pi/4$, with parameter values $c_1 = 1, c_2 = 2$.

**(S6)** Elliptical vortex in rotary mode (A16)

Similar to the hyperbolic case, elliptic solutions such as (A15) can rotate and deform at the same time. Here we examine the shape-preserving elliptical vortex (A16):

$$\psi = (c_1 \cos^2 Z + c_2 \sin^2 Z) x^2 + (c_1 \sin^2 Z + c_2 \cos^2 Z) y^2$$
$$\pm (c_1 - c_2) xy \sin 2Z, \quad c_1 \neq c_2,$$
$$p = 2c_1 c_2 (x^2 + y^2), \quad \rho = 0$$

As shown in Figure 6, this elliptical flow rotates by $\alpha = \pm Z(z)$ at each depth, and has constant eccentricity $e = \sqrt{1 - (c_1/c_2)^2}$.

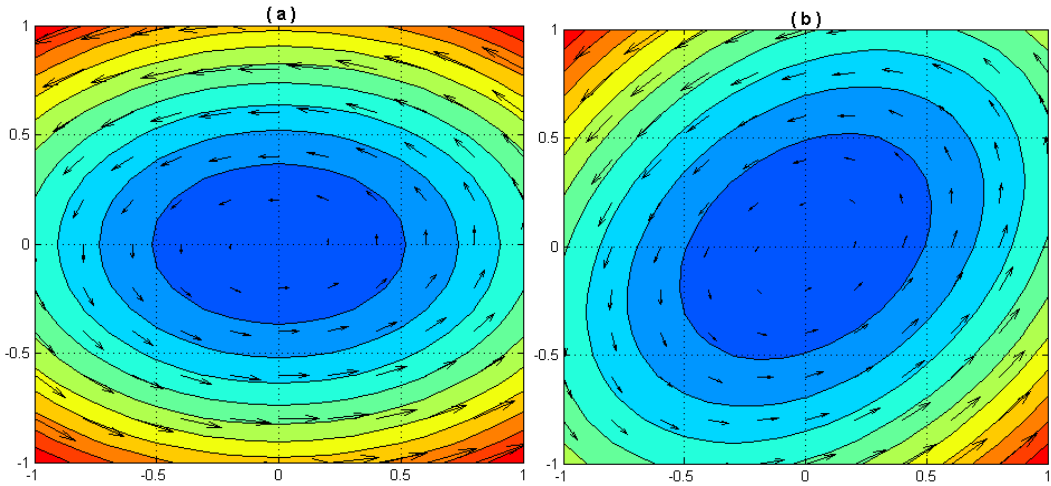

**Figure 6**. Streamfunction (color) and velocity vectors for solution (S6) at (a) $Z = 0$ and (b) $Z = \pi/4$, with parameter values $c_1 = 1, c_2 = 2$.

## 2) Rotating frame

Due to the additional f-terms, it is more difficult to obtain exact solutions for rotating frame. The following property, however, greatly reduces the complexity.

**Lemma 2 (Quadratic flow).** *In rotating frame a quadratic flow field does not rotate with depth.*

**Proof.** Adding $r_1 = 0$ and $r_2 = 0$ in (9) we have

$$(a_1 - a_3)a_2' - (a_1 - a_3)'a_2 = 0$$

which generates three cases after integration, i.e., $a_2 = 0$, $a_1 = a_3$ and $a_2 = c(a_1 - a_3)$. From (7), they correspond to $\alpha$ values of $0$, $\dfrac{\pi}{4}$ and $\dfrac{1}{2}\arctan(c)$ respectively. Therefore the rotation angle in regard to a Cartesian coordinate devoid of *xy* terms is always constant, meaning quadratic flows in rotating frame do not rotate with depth.

On one hand this lemma excludes the existence of rotary quadratic solution like (S5) in rotating frame. On the other hand it allows us to use coordinate rotation to eliminate the *xy* term and only look for solutions under $a_2 = 0$. The detailed solving process is given in Appendix B and the solutions are listed below.

**(S7)** Inertial circular vortex with skew center (B5)

$$\psi = -\frac{f}{2}[x + x_0(z)]^2 - \frac{f}{2}[y + y_0(z)]^2$$
$$p = 0, \quad \rho = 0$$

It describes free inertial motion in the absence of horizontal pressure gradients, with one-half a pendulum day as the period. The non-EB velocity field with skew center does not violate Theorem 2 because the centrifugal force here is balanced by the Coriolis force, leaving the pressure field unperturbed (Figure 7a).

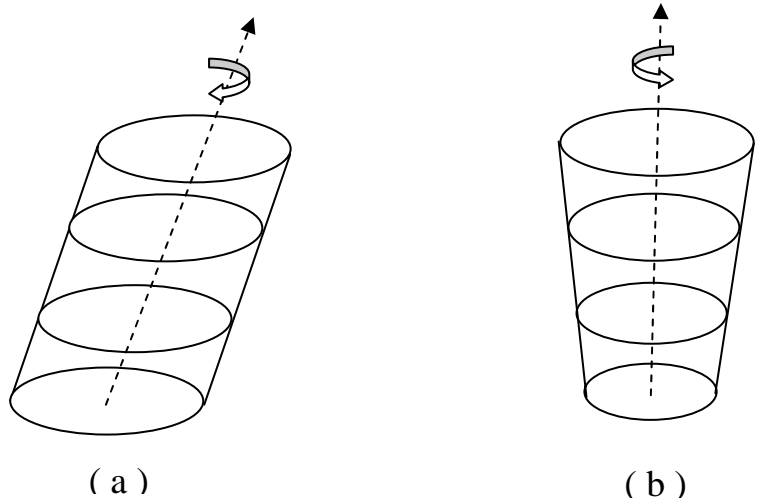

( a )            ( b )

**Figure 7**. (a) non-EB inertial circular vortex (S7). (b) EB circular vortex (S2 and S8).

**(S8)** Baroclinic circular vortex (B7)

$$\psi = a(z)(x^2 + y^2)$$
$$u = -2ay, \quad v = 2ax$$
$$p = a(2a + f)(x^2 + y^2)$$
$$\rho = -a'(4a + f)(x^2 + y^2)$$

The circular streamlines of this EB vortex are vertically aligned and concentric (Figure 7b).

**(S9)** Baroclinic straightline jet (B13)

$$\psi = [a(z)y + b(z)]^2$$
$$u = -2a(ay + b), \quad v = 0$$
$$p = f\psi, \quad \rho = -f\psi_z$$

In rotating fluid a straightline jet always incurs horizontal pressure variation and must be EB according to Theorem 2. Both circular vortex and straightline jet have constant speed along streamline and belong to constant-speed flow (CSF).

**(S10)** Inertial elliptic or hyperbolic flow (B6)

$$\psi = -\frac{f}{2}x^2 + c[y + y_0(z)]^2$$
$$u = -2c(y + y_0), \quad v = -f x$$
$$p = -\frac{f}{2}(2c + f) x^2, \quad \rho = 0, \quad c \neq -\frac{f}{2}$$

which describes a degenerate elliptic ($c<0$) or hyperbolic ($c>0$) flow with skew center.

**(S11)** Inertial parabolic flow (B17)

$$\psi = -\frac{f}{2}x^2 + a(z)y$$
$$u = -a(z), \quad v = -f x$$
$$p = -\frac{1}{2}f^2 x^2, \quad \rho = 0$$

Like elliptic and hyperbolic flows, the degenerate solution has variable speed along streamline and belongs to non-constant-speed flow (NCSF).

**(S12)** Baroclinic elliptical or hyperbolic flow (B12)

$$\psi = a(z)x^2 + b(z)y^2$$
$$u = -2by, \quad v = 2ax$$
$$p = 2ab(x^2 + y^2) + f\psi$$
$$\rho = -\frac{f(a-b)'}{a-b}\psi, \quad \zeta = 2(a+b)$$

where for elliptical flow,

$$b = \frac{c - f \operatorname{arctanh}(h)}{2h}, \quad a = bh^2$$

and for hyperbolical flow,

$$b = \frac{c - f \arctan(h)}{2h}, \quad a = -bh^2$$

The non-EB flow field pulsates with depth, similar to non-rotating solution S3 (see Figure 2 and Figure 3). Its symmetry axes do not rotate with depth, due to the requirement of Lemma 2.

**(S13)** Baroclinic parabolic flow (B18)

$$\psi = ax^2 + \frac{cay}{(f+2a)^2}$$

$$u = -\frac{ca}{(f+2a)^2}, \quad v = 2ax$$

$$p = fax^2 + \frac{cay}{f+2a}, \quad \rho = -\frac{fa'}{a}\psi$$

where $c$ is constant and $a = a(z)$. The streamfunction and velocity patterns are similar to solution S4 in Figure 4, except that in rotating frame the symmetry line of parabola is vertically aligned.

Table 1. Classification of quadratic solutions (CSF for constant-speed flow, NCSF for non-constant-speed flow, *italic* for baroclinic flows).

|  |  | Non-EB | EB |
|---|---|---|---|
| Non-rotating fluid | CSF | Straightline jet (**S1**) | Straightline jet (**S1**) <br> *Circular vortex* (**S2**) |
| | NCSF | Elliptical vortex (**S3, S6**) <br> Hyperbolic flow (**S3, S5**) <br> Parabolic flow (**S4**) | |
| Rotating Fluid | CSF | Inertial circular vortex with skew center (**S7**) | *Straightline jet* (**S8**) <br> *Circular vortex* (**S9**) |
| | NCSF | Inertial elliptic or hyperbolic flows (**S10**) <br> Inertial parabolic (**S11**) <br> *Elliptic vortex* (**S12**) <br> *Hyperbolic flow* (**S12**) <br> *Parabolic flow* (**S13**) | |

## IV. Discussion

The diverse quadratic flows are classified in Table 1 according to their geometric structure (EB or non-EB) and velocity property (CSF or NCSF). The rich topology of these quadratic solutions stands in contrast to the high-degree polynomial solutions obtained by *Sun* (2016). Unlike quadratic flows, the cubic and quartic solutions display reduced geometry in the form of straightline jet, circular vortex and multipolar strain field. The geometry reduction was explained by an analytical theorem stating that only straightline jet and circular vortex have functional solutions to the PIF model.

As an exact solution of the incompressible 2D Euler equations, Kirchhoff elliptic vortex in inertial frame rotates around the origin with constant angular velocity (*Lamb* 1932, Art. 159). The flow is steady if being viewed from a rotating frame. Meanwhile *Moore and Saffman* (1971) have obtained a 2D steady elliptic-vortex solution in uniform irrotational strain.

It is an interesting question how elliptic vortex looks like in 3D flows. Various pseudo-plane elliptic solutions from this study, including barotropic cases (S3, S6, S10) and baroclinic case (S12), all have non-aligned vertical structure. That is, their horizontal velocity vectors always change direction with depth. The simplest way to extend Kirchhoff elliptic vortex to 3D is by multiplying with a vertical structure

function $h(z)$ as

$$\psi = h(\frac{x^2}{a^2} + \frac{y^2}{b^2})$$
$$u = -\frac{2hy}{b^2}, \quad v = \frac{2hx}{a^2},$$

The flow is EB because its velocity vectors do not change direction with depth (here *a* and *b* are constant). The absence of this simple flow from the set of quadratic solutions is not surprising, as Theorem 1 dictates that streamlines of an EB pseudo-plane flow must be straightline or circular.

The geometric theory of pseudo-plane flows also sheds light on steady-state solutions to traditional shallow-water reduced-gravity models, which have been commonly used in describing oceanic vortices. *Ruddick* (1987) found a steady solution of elliptical ring in strain flow resembling the two-dimensional solutions of *Moore and Saffman* (1971) and *Kida* (1981). But his result appears to contradict *Cushman-Roisin et al.* (1985) which showed a simple reduced-gravity model only admits straightline jet and circular vortex as equilibrium solutions (a detailed derivation is provided in Appendix C). It also appears to violate Theorem 1 if we notice that layered reduced gravity models are fundamentally baroclinic and can be regarded as vertically integrated pseudo-plane flows in each layer.

The paradox is resolved with the knowledge that the two-layer model in *Cushman-Roisin et al.* (1985) is effectively EB since the lower layer has no motion, but *Ruddick*'s three-layer model is non-EB because the prescribed strain flows in the upper and bottom layers differ from the elliptic flow in the middle layer. Both EB circular vortex and non-EB elliptic vortex are in agreement with Theorem 1.

## V. Conclusion

As an effort to extend quadratic-flow study to three-dimensional stratified fluid, the study obtains the complete set of exact quadratic solutions to the PIF model. Key geometric results include: the existence of non-EB baroclinic solutions disproves an earlier conjecture by Sun08; only straightline jet and circular vortex are found to have EB structure, in agreement with Theorem 1; quadratic flows in rotating frame do not rotate with depth. The rich topology of quadratic pseudo-plane flows is in contrast to the reduced geometry of high-degree polynomial solutions, validating the analytical theory proposed by *Sun* (2016).

**ACKNOWLEDGEMENTS**

This study was supported by the National Natural Science Foundation of China (Grant No. 41576017).

# APPENDIX A: Derivation of non-rotating quadratic solutions

For non-rotating flows, $f = 0$, coefficients in (9) are reduced to

$$\begin{aligned} &r_1 = -a_2 C, \quad r_2 = a_2 C, \quad r_3 = (a_1 - a_3)C \\ &r_4 = 2a_1 A + a_2 B - 2b_2 C, \quad r_5 = a_2 A + 2a_3 B + 2b_1 C \\ &r_6 = b_1 A + b_2 B \end{aligned} \quad (A1)$$

There are two cases according to $a_2$ value.

**Case 1.** $a_2 = 0$.

After dropping constant coefficients, (A1) is reduced to

$$\begin{aligned} r_3 &= (a_1 - a_3)(a_1 a_3)' \\ r_4 &= a_1(a_1 b_2)' - b_2(a_1 a_3)' \\ r_5 &= -a_3(a_3 b_1)' + b_1(a_1 a_3)' \\ r_6 &= b_1(a_1 b_2)' - b_2(a_3 b_1)' \end{aligned} \quad (A2)$$

in which $r_3 = 0$ requires $a_1 a_3 = const.$ or $a_1 = a_3$. For both cases, $\Delta = 4a_1 a_3$ does not change sign with depth.

**Case 1.1.** $a_1 = a_3 \neq 0$

From $r_4 = r_5 = 0$ in (A2) we have

$$a_1 b_1' - a_1' b_1 = 0, \quad a_1 b_2' - a_1' b_2 = 0,$$

integration of which gives $b_1 = a_1(z)c_1$, $b_2 = a_1(z)c_2$, where $c_1, c_2$ are constant. The solution is $\psi = a_1(z)(x^2 + y^2 + c_1 x + c_2 y)$ and can be centralized by horizontal translation as

$$\psi = a_1(z)(x^2 + y^2) \quad (A3)$$

**Case 1.2,** $a_1 a_3 = const.$

For a quadratic flow we assume $a_1 \neq 0$ and $a_3 = c/a_1$, where $c$ is a constant. Then $r_4 = r_5 = 0$ in (A2) gives $(a_1 b_2)' = 0$ and $c(b_1/a_1)' = 0$.

If $a_3 = 0$, $c = 0$, $a_1 b_2 = c_2$, the solution is a parabolic flow

$$\psi = a_1(z)x^2 + b_1(z)x + \frac{c_2}{a_1(z)} y \qquad (A4)$$

If $a_3 \neq 0$, $c \neq 0$, we have $b_1 = 2c_1 a_1$, $a_1 b_2 = 2cc_2$ and

$$\psi = a_1(x+c_1)^2 + \frac{c}{a_1}(y+c_2)^2$$

The solution is elliptic or hyperbolic flow. It has fixed center and can be centralized as

$$\psi = a_1 x^2 + \frac{c}{a_1} y^2 \qquad (A5)$$

**Case 2.** $a_2 \neq 0$

$r_2 = 0$ in (A1) requires $C = 0$, integration of which yield a constant $\Delta$. Combining with the case of $a_2 = 0$ at (A2), we prove Lemma 1 in Section 3.1.

Now (A1) becomes

$$\begin{aligned} r_4 &= 2a_1 A + a_2 B \\ r_5 &= a_2 A + 2a_3 B \\ r_6 &= b_1 A + b_2 B \end{aligned} \qquad (A6)$$

There are three scenarios according to $A$ and $B$:

1) One of $A$ and $B$ is zero, say $A = 0, B \neq 0$. Then we have $a_2 = a_3 = b_2 = 0$. The solution is an EB straightline jet $\psi = a_1 x^2 + b_1 x$.

2) Both $A$ and $B$ are non-zero. Then $r_5 = r_6 = 0$ gives $a_2 b_2 = 2a_3 b_1$ and consequently $B = 0$, which is contradictory.

3) Both $A$ and $B$ are zero. Then

$$2a_1b_2 - a_2b_1 = c_1$$
$$a_2b_2 - 2a_3b_1 = c_2 \quad (A7)$$

both are constant. The following are two cases according to $\Delta$ value.

**Case 2.1**. $\Delta = 0$, $a_2 = \pm 2\sqrt{a_1 a_3}$. Then (A7) gives,

$$b_2 = (a_2 b_1 + c_1)/2a_1, \quad c_1\sqrt{a_3} = \pm c_2\sqrt{a_1} \quad (A8)$$

Substituting (A8) into $\psi$ we have

$$\begin{aligned}\psi &= a_1 x^2 \pm 2\sqrt{a_1 a_3}\, xy + a_3 y^2 + b_1 x + (a_2 b_1 + c_1) y / 2a_1 \\ &= (\sqrt{a_1} x \pm \sqrt{a_3} y)^2 + (\sqrt{a_1} x \pm \sqrt{a_3} y) b_1 / \sqrt{a_1} + c_1 y / 2a_1 \\ &= (\sqrt{a_1} x \pm \sqrt{a_3} y + b_1 / 2\sqrt{a_1})^2 + c_1 y / 2a_1 \end{aligned} \quad (A9)$$

If $c_1 \neq 0$, (A8) gives $\sqrt{a_3} = \pm\sqrt{a_1}\, c_2/c_1 = c\sqrt{a_1}$ and (A9) yields a non-EB parabolic solution

$$\psi = a_1(x + cy + \frac{b_1}{2a_1})^2 + \frac{c_1}{2a_1} y \quad (A10)$$

Clearly, coordinate rotation can eliminate the $xy$ term in (A10) and convert it to (A4).

If $c_1 = 0$, (A9) yields a veering straightline jet

$$\psi = (\sqrt{a_1} x \pm \sqrt{a_3} y + \frac{b_1}{2\sqrt{a_1}})^2 \quad (A11)$$

which is not included in (A10) by taking $c_1 = 0$ there.

**Case 2.2**. $\Delta = c \neq 0$. Then (A7) has a unique solution

$$b_1 = 2a_1 c_1 + a_2 c_2, \quad b_2 = a_2 c_1 + 2a_3 c_2$$

Substituting it into $\psi$ gives

$$\psi = a_1(x + c_1)^2 + a_2(x + c_1)(y + c_2) + a_3(y + c_2)^2$$

The flow is vertically concentric and can be moved to the origin by horizontal translation, which yields a centralized general solution

$$\psi = a_1 x^2 \pm xy\sqrt{4a_1 a_3 - c} + a_3 y^2$$
$$p = \frac{1}{2} c(x^2 + y^2), \quad \rho = 0 \tag{A12}$$

If $\Delta = c < 0$, (A12) is hyperbolic flow and can be represented by hyperbolic functions as $a_1 = c_1 \sinh Z$, $a_3 = c_2 \sinh Z$, $a_2^2 = 4c_1 c_2 \cosh^2 Z$, where $c_1$ and $c_2$ are positive constants. It gives

$$\psi = (c_1 x^2 + c_2 y^2)\sinh Z \pm 2\sqrt{c_1 c_2}\, xy \cosh Z \tag{A13}$$

The flow can also be represented by circular functions to yield a shape-preserving asymmetric strain flow that rotates with depth

$$\psi = (c_1 \cos^2 Z - c_2 \sin^2 Z) x^2 + (c_1 \sin^2 Z - c_2 \cos^2 Z) y^2 \\ \pm (c_1 + c_2) xy \sin 2Z \tag{A14}$$

If $\Delta = c > 0$, (A12) is elliptical flow and can be represented by hyperbolic functions as $a_1 = c_1 \cosh Z$, $a_3 = c_2 \cosh Z$, $a_2^2 = 4c_1 c_2 \cosh^2 Z$. It gives

$$\psi = (c_1 x^2 + c_2 y^2)\cosh Z \pm 2\sqrt{c_1 c_2}\, xy \sinh Z \tag{A15}$$

The flow can also be expressed by circular functions like

$$\psi = (c_1 \cos^2 Z + c_2 \sin^2 Z) x^2 + (c_1 \sin^2 Z + c_2 \cos^2 Z) y^2 \\ \pm (c_1 - c_2) xy \sin 2Z \tag{A16}$$

which has fixed major-to-minor axis ratio at each depth.

# APPENDIX B: Derivation of rotating quadratic solutions

Lemma 2 in Section 3.2 states that in rotating frame quadratic flows do not rotate with depth. As a result the $xy$ term can be removed by a coordinate rotation. We only need to consider the $a_2 = 0$ scenario.

**Case 1.** Elliptical and hyperbolic Solutions ($a_2 = 0$, $a_1 \neq 0$, $a_3 \neq 0$).

Streamfunction (6) is rewritten as

$$\psi = a_1(x+\frac{b_1}{2a_1})^2 + a_3(y+\frac{b_2}{2a_3})^2 = a_1(x+x_0)^2 + a_3(y+y_0)^2 \tag{B1}$$

$r_3 = 0$ in (22) gives

$$2(a_1 - a_3)(a_1 a_3)' + (a_1 a_3' - a_3 a_1')f = 0 \tag{B2}$$

With $b_1 = 2a_1 x_0$, $b_2 = 2a_3 y_0$ and (B2), we can simplify $r_4 = 0$ and $r_5 = 0$ to

$$(2a_1 + f) y_0' = 0 \tag{B3}$$

$$(2a_3 + f) x_0' = 0 \tag{B4}$$

There are three cases according to $a_1$ and $a_3$.

**Case 1.1.** Both $a_1$ and $a_3$ equal $-\frac{f}{2}$.

Then (B2-B4) are satisfied, and the flow is inertial circular vortex

$$\psi = -\frac{f}{2}[(x+x_0)^2 + (y+y_0)^2] \tag{B5}$$

**Case 1.2.** Only one of $a_1$ and $a_3$ equals $-\frac{f}{2}$.

Let's say $a_1 = -\frac{f}{2}$, $a_3 \neq -\frac{f}{2}$. Then (B2) and (B4) require $x_0$ and $a_3$ to be constant. The solution is a partially inertial flow

$$\psi = -\frac{f}{2}x^2 + c(y+y_0)^2 \tag{B6}$$

**Case 1.3**. Neither $a_1$ or $a_3$ equals $-\dfrac{f}{2}$.

Then (B3) and (B4) require $x_0$ and $y_0$ to be constant.

If $a_1 = a_3$, we obtain a concentric circular solution

$$\psi = a_1(z)(x^2 + y^2) \tag{B7}$$

If $a_1 \neq a_3$, first transform (B2) into

$$2(a_3 - a_1)(a_1 a_3)' + f\, a_3^2 \left(\dfrac{a_1}{a_3}\right)' = 0 \tag{B8}$$

To facilitate integration, assume

$$a_1 = \pm a_3\, h^2(z) \tag{B9}$$

where the sign correspond to elliptical or hyperbolical case. Substituting it into (B8), we have

$$2(a_3 h)' + \dfrac{f\, h'}{1 \mp h^2} = 0,$$

which can be integrated by using the definition of inverse hyperbolic tangent function

$$\operatorname{arctanh}(z) = \int_0^z \dfrac{dt}{1 - t^2}$$

and inverse tangent function

$$\arctan(z) = \int_0^z \dfrac{dt}{1 + t^2}.$$

Correspondingly it yields an elliptical solution

$$a_3 = \dfrac{c - f\, \operatorname{arctanh}(h)}{2h}, \quad a_1 = a_3 h^2 \tag{B10}$$

and a hyperbolical solution

$$a_3 = \dfrac{c - f\, \arctan(h)}{2h}, \quad a_1 = -a_3 h^2 \tag{B11}$$

The full solutions for both cases have the same form

$$\psi = a_1 x^2 + a_3 y^2$$
$$u = -2a_3 y, \quad v = 2a_1 x$$
$$p = 2a_1 a_3 (x^2 + y^2) + f\psi \quad \text{(B12)}$$
$$\rho = -\frac{f(a_1 - a_3)'}{a_1 - a_3}\psi$$

In density we have used (B8) to eliminate the $(a_1 a_3)'$ term. Similar solutions have been obtained by *Jia et al.* (2012).

**Case 2.** Parabolic solution ($a_2 = a_3 = 0$).

Streamfunction (6) becomes

$$\psi = a_1 x^2 + b_1 x + b_2 y$$

which is rewritten as

$$\psi = a_1(x + x_0)^2 + b_2 y, \quad x_0 = \frac{b_1}{2a_1}$$

If $b_2 = 0$, the solution is a straightline jet

$$\psi = a_1(x + x_0)^2 \quad \text{(B13)}$$

If $b_2 \neq 0$, the flow is parabolic. Most coefficients in (22) drop out and $r_4 = r_6 = 0$ gives

$$2a_1(a_1 b_2)' + (a_1 b_2' - a_1' b_2)f = 0 \quad \text{(B14)}$$

$$2b_1(a_1 b_2)' + (b_1 b_2' - b_1' b_2)f = 0 \quad \text{(B15)}$$

Substitution of (B14) and $b_1 = 2a_1 x_0$ into (B15) yields

$$f a_1 b_2 x_0' = 0$$

It requires $x_0$ to be constant, which means the vertex points of parabola streamlines align vertically. We apply horizontal translation to shift the vertex to the origin and assume $b_1 = 0$ hereafter. With (B15) becoming redundant, rewrite (B14) as

$$(2a_1 + f)a_1 b_2' + (2a_1 - f)a_1' b_2 = 0 \tag{B16}$$

There are two cases according to $a_1$.

**Case 2.1.** $a_1 = -\dfrac{f}{2}$.

(B16) is satisfied and the solution is

$$\psi = -\frac{f}{2}x^2 + b_2 y \tag{B17}$$

**Case 2.2.** $a_1 \neq -\dfrac{f}{2}$.

Assuming $b_2 = a_1 h(z)$, we rewrite (B16) as

$$\frac{h'}{h} = -\frac{4a_1'}{f + 2a_1},$$

the integration of which yields

$$h(z) = \frac{c}{(f + 2a_1)^2}$$

The solution is

$$\psi = a_1 x^2 + \frac{c a_1 y}{(f + 2a_1)^2} \tag{B18}$$

## APPENDIX C: Steady solutions to the shallow-water equations

The two-layer shallow-water reduced gravity model, such as used in *Cushman-Roisin et al.* (1985), consists of one moving layer above a motionless layer. Its governing equations are

$$\frac{\partial u}{\partial t} + u\frac{\partial u}{\partial x} + v\frac{\partial u}{\partial y} - fv = -g'\frac{\partial h}{\partial x},$$
$$\frac{\partial v}{\partial t} + u\frac{\partial v}{\partial x} + v\frac{\partial v}{\partial y} + fu = -g'\frac{\partial h}{\partial y}, \quad \text{(C1)}$$
$$\frac{\partial h}{\partial t} + \frac{\partial (hu)}{\partial x} + \frac{\partial (hv)}{\partial y} = 0,$$

where $g' = g(\rho_2 - \rho_1)/\rho_0$ is the reduced gravity and $h$ is the depth of upper layer. Except for the replacement of gravity $g$ by $g'$ and the full water depth by $h$, this reduced gravity model is identical to the shallow-water equations over a flat bottom as used in *Young* (1986).

Both shallow water model and two-layer reduced gravity model are essentially EB (one is barotropic EB and one is baroclinic EB). The steady state of both models can be regarded as a vertically integrated pseudo-plane two-dimensional flow.

Assume thickness $h$ is a quadratic function and velocities are linear functions of spatial coordinates. (For such EB models, the $xy$ term can always be eliminated by rotating the coordinates.)

$$g'h = Ax^2 + Cy^2 + Dx + Ey + F$$
$$u = U_1 x + U_2 y + U_0, \quad v = V_1 x + V_2 y + V_0 \quad \text{(C2)}$$

All coefficients in (C2) are constant. For quadratic flow, assume $A \neq 0$. Substituting into (C1) and requiring the coefficients of the resultant ODE to be zero everywhere would yield 12 algebraic equations:

$$U_1U_2 + (U_2 - f)V_2 = 0$$
$$V_1V_2 + (V_1 + f)U_1 = 0$$
$$2A + U_1^2 + (U_2 - f)V_1 = 0$$
$$2C + V_2^2 + (V_1 + f)U_2 = 0$$
$$(3U_1 + V_2)A = 0$$
$$(U_1 + 3V_2)C = 0 \quad \text{(C3)}$$
$$U_2A + V_1C = 0$$
$$D + U_0U_1 + (U_2 - f)V_0 = 0$$
$$E + V_0V_2 + (V_1 + f)U_0 = 0$$
$$2U_0A + (2U_1 + V_2)D + V_1E = 0$$
$$2V_0C + U_2D + (U_1 + 2V_2)E = 0$$
$$U_0D + V_0E + (U_1 + V_2)F = 0$$

(C3.1) and (C3.2) require

$$(U_1 + V_2)(U_2 - V_1 - f) = 0$$

which has two cases.

**Case 1**, $V_2 = -U_1$.

From (C3.1) we have $U_1 = 0$. Therefore $V_2 = 0$. Then (C3.3) and (C3.4) become $2A = (f - U_2)V_1$, $2C = -(V_1 + f)U_2$. Substituting them into (C3.7) gives $V_1 = -U_2$, $A = C$ or $U_2 = 0, C = 0$.

It leads to the following two cases.

**Case 1.1**, $V_1 = -U_2$, $A = C = (U_2 - f)U_2/2$, $U_1 = V_2 = 0$.

From (C3.8) and (C3.9) we have $D = (f - U_2)V_0$, $E = (U_2 - f)U_0$.

The solution is a circular vortex,

$$g'h = \frac{1}{2}(U_2 - f)U_2(x^2 + y^2) + (U_2 - f)(U_0y - V_0x) + F$$
$$u = U_2y + U_0, \quad v = -U_2x + V_0$$

The vortex can be centralized to the origin by letting $U_0 = V_0 = 0$

$$g'h = \frac{1}{2}a(a-f)(x^2+y^2) \tag{C4}$$
$$u = ay, \quad v = -ax$$

which is essentially the solution (17) in *Cushman-Roisin et al.* (1985).

**Case 1.2**, $U_2 = 0, C = 0, 2A = fV_1, U_1 = V_2 = 0$

From (C3.8) and (C3.9) we have $D = fV_0, E = U_0 = 0$.

The solution is a straightline jet

$$g'h = \frac{1}{2}fV_1x^2 + fV_0x + F \tag{C5}$$
$$u = 0, \quad v = V_1x + V_0$$

and satisfies geostrophic balance $fv = g'h_x$.

**Case 2**, $V_1 = U_2 - f$.

Since $A \neq 0$, (C3.5) requires $V_2 = -3U_1$, and (C3.6) reduces to $U_1C = 0$.

If $U_1 \neq 0, C = 0$, (C3.7) gives $U_2 = 0$ and (C3.4) gives $U_2^2 + 9U_1^2 = 0$, leading to $U_1 = 0$. It is contradictory.

Then $U_1 = 0$, (C3.1) gives $V_1V_2 = 0$ and (C3.3) gives $2A = -V_1^2 \neq 0$, which requires $V_2 = 0$. (C3.4) reduces to $2C = -U_2^2$ and (C3.7) gives

$$U_2V_1(U_2 + V_1) = U_2V_1(2U_2 - f) = 0$$

If $U_2 = 0$, we obtain

$$2A = -f^2, C = 0, U_1 = 0, U_2 = 0, V_1 = -f, V_2 = 0, D = fV_0, E = 0, U_0 = 0, V_0 = 0$$

which is a special case of straightline solution (C5):

$$g'h = -\frac{1}{2}f^2x^2 + fV_0x + F \tag{C6}$$
$$u = 0, \quad v = -fx + V_0$$

If $2U_2 = f$, we have

$$A = -\frac{f^2}{8}, C = -\frac{f^2}{8}, U_1 = 0, U_2 = \frac{f}{2}, V_1 = -\frac{f}{2}, V_2 = 0, D = \frac{1}{2}fV_0, E = -\frac{1}{2}fU_0$$

The solution is a special case of circular-vortex solution (C4) with inertial period:

$$g'h = -\frac{1}{8}f^2(x^2 + y^2) + \frac{1}{2}fV_0 x - \frac{1}{2}fU_0 y + F$$
$$u = \frac{1}{2}fy + U_0, \quad v = -\frac{1}{2}fx + V_0$$
(C7)

We thus prove that steady quadratic solutions of the shallow-water equations and two-layer reduced-gravity model are either straightline jet or circular vortex.